\documentclass[aps,prl,twocolumn,preprintnumbers,amsmath,amssymb]{revtex4}
\usepackage{graphicx}
\usepackage{bm}
\usepackage[final]{pdfpages}

\begin{document}
\title{Large linear magnetoresistance in Dirac semi-metal Cd$_3$As$_2$ with Fermi surfaces close to the Dirac points}
\date{\today}

\author{Junya Feng}
\author{Yuan Pang}
\author{Desheng Wu}
\author{Zhijun Wang}
\author{Hongming Weng}
\author{Jianqi Li}
\author{Xi Dai}
\author{Zhong Fang}
\author{Youguo Shi}\email[Corresponding author for crystal growth: ]{ygshi@iphy.ac.cn}
\author{Li Lu}\email[Corresponding author: ]{lilu@iphy.ac.cn}
\affiliation{Beijing National Laboratory for Condensed Matter Physics, Institute of Physics, Chinese Academy of Sciences\\
Collaborative Innovation Center of Quantum Matter, Beijing 100190, People's Republic of China}

\begin{abstract}
We have investigated the magnetoresistive behavior of Dirac semi-metal Cd$_3$As$_2$ down to low temperatures and in high magnetic fields. A positive and linear magnetoresistance (LMR) as large as 3100\% is observed in a magnetic field of 14 T, on high-quality single crystals of Cd$_3$As$_2$ with ultra-low electron density and large Lande $g$ factor. Such a large LMR occurs when the magnetic field is applied perpendicular to both the current and the (100) surface, and when the temperature is low such that the thermal energy is smaller than the Zeeman splitting energy. Tilting the magnetic field or raising the temperature all degrade the LMR, leading to a less pronounced quadratic behavior. We propose that the phenomenon of LMR is related to the peculiar field-induced shifting/distortion of the helical electrons' Fermi surfaces in momentum space.
\end{abstract}

\maketitle

Compared with those negative magnetoresistive behaviors such as giant magnetoresistance \cite{GMR} and colossal magnetoresistance \cite{CMR} whose mechanisms have been well understood, positive large LMR was also reported in past decades but its mechanism is not fully clarified. Such behavior was found in highly disordered nonmagnetic narrow-band semiconductors such as Ag$_{2\pm\delta}$Te and Ag$_{2\pm\delta}$Se \cite{Xu_R}, in bismuth thin films \cite{LMRBi}, and in Dirac electron systems such as epitaxial graphene \cite{LMRGraphene} and topological insulators-related materials \cite{LMRBiTe,LMRBiSe,Xue_Chen,SYLi,cdasOng}. Large LMR was also observed in InSb \cite{LMRInSb}, a material with very small electron effective mass and very large electron Lande $g$ factor. Several theories have been proposed to explain the phenomenon. Abrikosov proposed that the LMR is a quantum magnetoresistance of linearly dispersed electron systems, arising when all the electrons are filled in the first Landau level (LL), i.e., in the extreme quantum limit \cite{Abrikosov98,Abrikosov2000}. Wang and Lei proposed that the LMR can still arise when the LLs are smeared, if with a positive $g$ factor \cite{WangandLei}. There are also pictures involving no LLs. Parish and Littlewood explained the LMR in Ag$_{2\pm\delta}$Te and Ag$_{2\pm\delta}$Se by modeling the materials as a network due to disorder-induced mobility fluctuation \cite{Littlewood03}. So far the mechanism of LMR is still waiting to be clarified.

In this work, we revisit the LMR issue by investigating the magnetoresistive behavior of Cd$_3$As$_2$ single crystals. Cd$_3$As$_2$ is predicted to be a three-dimensional (3D) Dirac semimetal \cite{FZCd3As2}, with linearly dispersed electron states in the bulk, and Fermi arcs at the surface which connect the bulk Dirac cones. The existence of 3D Dirac cones has been confirmed by angular resolved photoemission spectroscopy study \cite{3DDCHasan}. And a LMR behavior has recently been observed on samples with Fermi level well above the Dirac point \cite{SYLi,cdasOng}, i.e., with carrier density of the order 10$^{18}$ cm$^{-3}$. Here, we report our investigations on single crystals of Cd$_{\rm 3}$As$_{\rm 2}$ with a much lower carrier density, such that the Fermi surfaces are small spheres very close to the Dirac points in the momentum space, rather than near the Lifshitz transition (i.e., touching with each other). We observed a positive, very large and direction-dependent LMR up to 3100\% in a field of 14 T at 2 K.

\begin{figure}
\includegraphics[width=0.85 \linewidth]{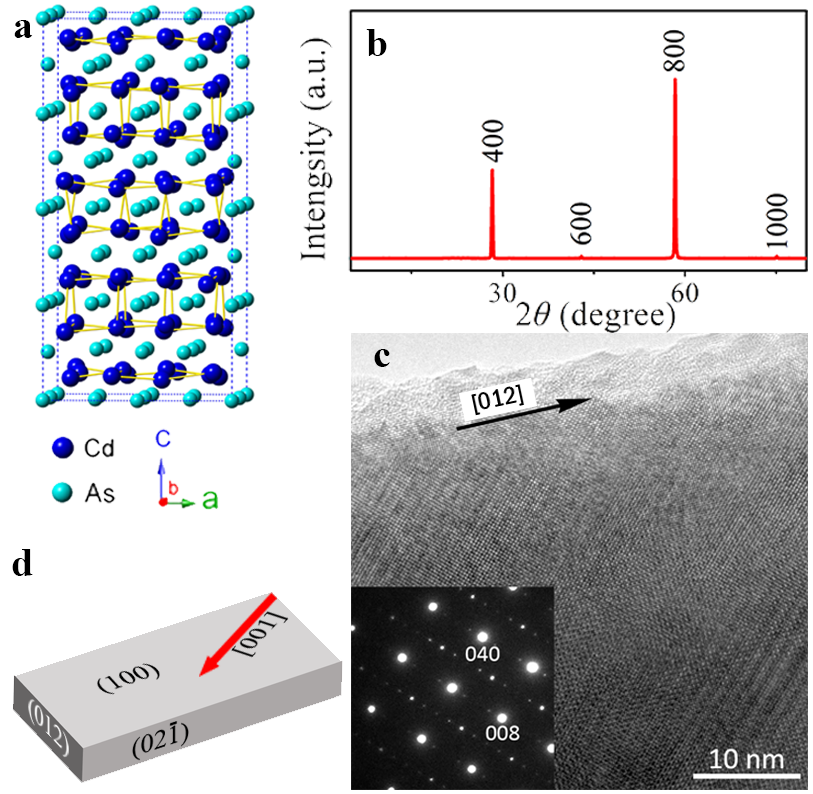}
\caption{\label{fig:fig1} {(color online) \textbf{a}, Crystal structure of Cd$_3$As$_2$ with I$_{41}$/acd centrosymmetry viewed along the [010] direction (the b-direction). \textbf{b}, X-ray diffraction pattern of plate-like rectangular Cd$_3$As$_2$ crystals used in this experiment. \textbf{c}, High resolution transmission electron microscope image and selected-area electron diffraction pattern (inset) of a plate-like Cd$_3$As$_2$ crystal taken along the [100] zone-axis direction, i.e., the electron beam incidents perpendicularly to the largest facet of the crystal. \textbf{d}, The indexes of the facets of the plate-like crystals. The arrow indicates the [001] zone-axis direction.}}
\end{figure}

Single crystals of Cd$_3$As$_2$ were synthesized via a two-step chemical vapor transport method according to the details presented in the supplementary material. Two different phases of Cd$_3$As$_2$ were obtained, one with P$_{42}$/nmc symmetry\cite{symmetryP42nmc} and the other with I$_{41}$/cd\cite{symmetryI41cd} or I$_{41}$/acd\cite{symmetryI41acd} symmetry. The plate-like rectangular crystals studied in this experiment belong to the latter phase, whose crystal structure, X-ray diffraction (XRD) pattern and high-resolution transmission electron microscope (HRTEM) image are shown in Figs. 1\textbf{a}, 1\textbf{b} and 1\textbf{c}, respectively. All the peaks in Fig. 1\textbf{b} can be well indexed to the centered tetragonal structure of I$_{41}$/acd symmetry, with lattice parameters $a$=$b$=12.6527{\AA} and $c$=25.4578{\AA}. HRTEM study indicates that the plate-like crystals grow preferentially along the [012] direction (the length direction). The width direction is along [02$\bar{1}$]. The largest facet of the crystal is the (100) plane.

Hall-bar devices were defined by manually attaching six Pt wires of diameter 20 $\mu$m to each thin-plate crystal with the use of silver paste. The electric current was applied along the length direction, and the magnetic field was applied in the plane defined by the thickness direction and the width direction, as illustrated in the inset of Fig. 4\textbf{b}. The longitudinal and Hall resistance measurements were carried out using a standard lock-in technique in a PPMS system (Quantum Design) which has a base temperature of 2 K and a magnetic field up to 14 T.

About ten Hall-bar devices were studied, while some of the devices yielded unphysical negative resistance --- which were presumably due to non-uniform current flowing caused by the existence of micro-cracks in the crystals, reliable and reproducible results were finally obtainable. Presented here are the data measured on one of the Hall-bar devices whose optical image is shown in the inset of Fig. 2\textbf{a}. For this device, the width of the crystal is $\sim$ 150 $\rm\mu m$, the distance between the voltage contacts is $\sim$ 450 $\rm\mu m$, and the thickness of the crystal is $\sim$ 20 $\rm\mu m$.

\begin{figure}
\includegraphics[width=0.99 \linewidth]{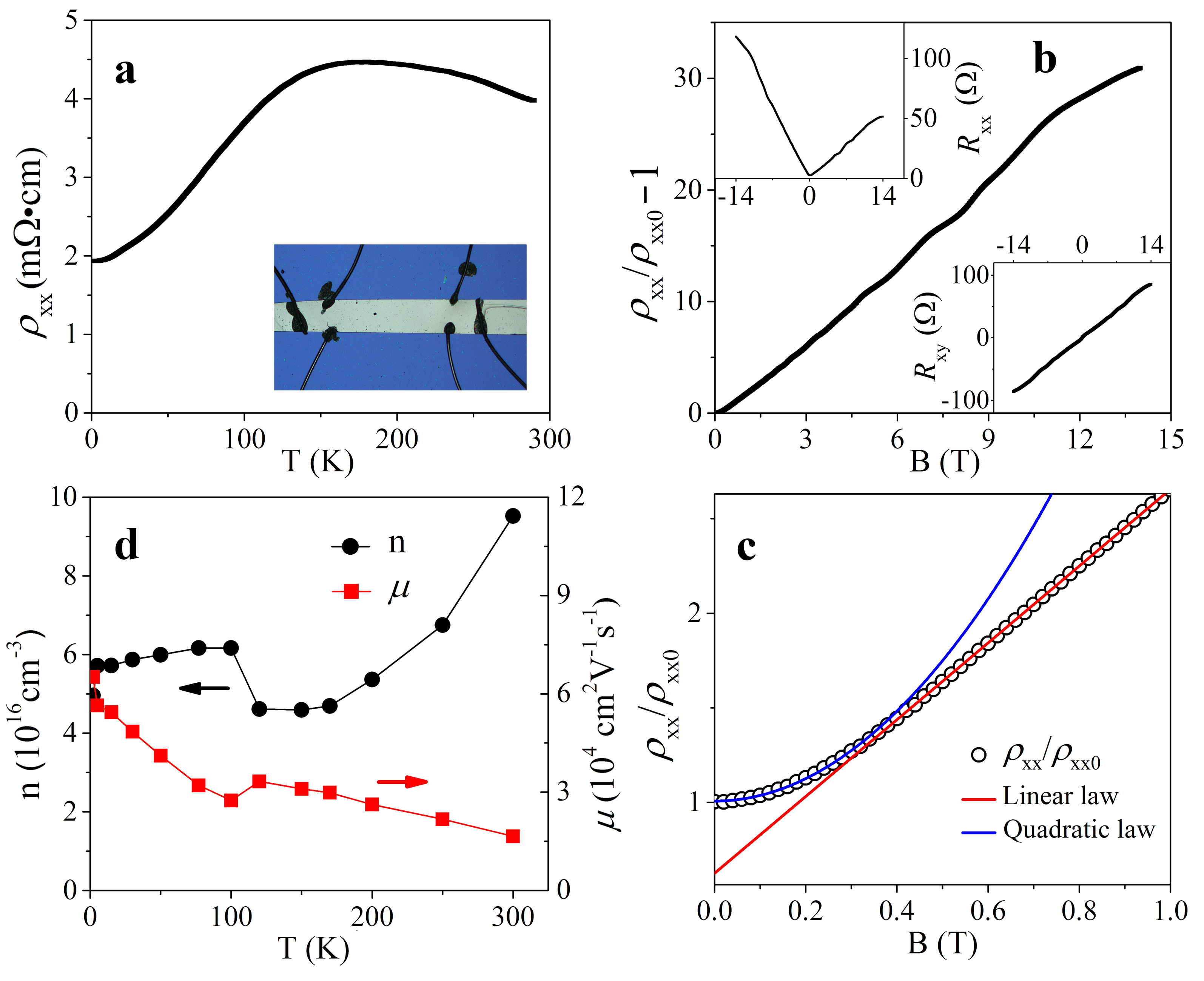}
\caption{\label{fig:fig2} {(color online) \textbf{a}, The temperature dependence of longitudinal resistivity $\rho_{\rm xx}$ of Cd$_3$As$_2$. Inset: the optical image of the Hall-bar device defined on a thin-plate of Cd$_3$As$_2$ crystal, on which the data shown here were measured. \textbf{b}, Main frame: the magneto-resistivity $\rho_{\rm xx}/\rho_{\rm xx0}-1$ measured at 2 K (where $\rho_{\rm xx0}$ is the longitudinal resistivity in zero magnetic field). These data have been symmetrized by averaging over the positive and negative field directions. Upper left inset: original (un-symmetrized) data of $R\rm _{xx}$. Lower right inset: the Hall resistivity $R_{\rm xy}$ measured at 2 K. \textbf{c}, Details of $\rho_{\rm xx}/\rho_{\rm xx0}$ below $B$=1 T. \textbf{d}, The temperature dependencies of the carrier density and the mobility (deduced from the Hall coefficient using simple Drude mode).}}
\end{figure}

Figure 2\textbf{a} shows the resistivity of the crystal as a function of temperature. It has a semiconductor-like temperature dependence near room temperature and a metallic behavior below $\sim$100 K. We note that the high-temperature semiconducting behavior was only observable on crystals with very low carrier density. The phenomenon was reproducible on crystals of different batches. This semiconducting behavior is presumably contributed by the thermally excited carriers due to the existence of, e.g., point defects in the crystal, similar to the case in InSb \cite{LMRInSb}. With decreasing temperature, the thermally excited carriers gradually freeze out, leading to a less temperature-dependent carrier density (Fig. 2\textbf{d}) and thus a metallic temperature dependence of resistivity below $\sim$100 K.

Figure 2\textbf{b} shows the field dependencies of the longitudinal and the Hall resistance of the crystal at 2 K. The measured original data of the former and the latter are shown in the upper left and the lower right insets, respectively. Because the Hall signal could be much larger than the longitudinal one, it could be easily picked up in the $\rho_{\rm xx}$ measurement. Nevertheless, the Hall signal can be removed by averaging the $\rho_{\rm xx}$ data over positive and negative field directions. The symmetrized data, represented as $\rho_{\rm xx}/\rho_{\rm xx0}-1$ (where $\rho_{\rm xx0}$ denotes the longitudinal resistance in zero magnetic field), are shown in the main frame of Fig. 2\textbf{b}. A very large LMR up to 3100\% in a field of 14 T can be seen, accompanied with noticeable Shubnikov-de Haas (SdH) oscillations. Figure 2\textbf{c} further shows that the LMR behavior extends to a field as low as $\sim$0.35 T, where no sign of SdH oscillations can be resolved. Below $\sim$0.35 T a quadratic law dominates.

From the Hall slope $R_{\rm H}=-1/ne$, the carrier density can be deduced: $n=5.0\times 10^{16} \rm cm^{-3}$. According to the Drude conductivity $\sigma=ne\mu$, the effective electron mobility is then: $\mu=6.5\times10^{4}{\rm cm^{2}V^{-1}s^{-1}}$. The temperature dependencies of these two quantities are plotted in Fig. 2\textbf{d}. It can be seen that $n$ is almost constant below $\sim$100 K, whereas $\mu$ changes by a factor of 2. The $n$ estimated from the Hall measurement roughly agrees with that estimated from the SdH oscillation after ellipsoid correction, which is $1.25\times 10^{17} \rm cm^{-3}$, indicating that the measured signals come from a 3D electron system in Cd$_{\rm 3}$As$_{\rm 2}$. The estimation of the carrier density from the SdH oscillations, together with the estimations of the effective electron mass $m^{\ast}=0.025 m_{\rm e}$ and the Dingle temperature $T_{\rm D}$=12 K, can be found in the supplementary material.

\begin{figure}
\includegraphics[width=0.99 \linewidth]{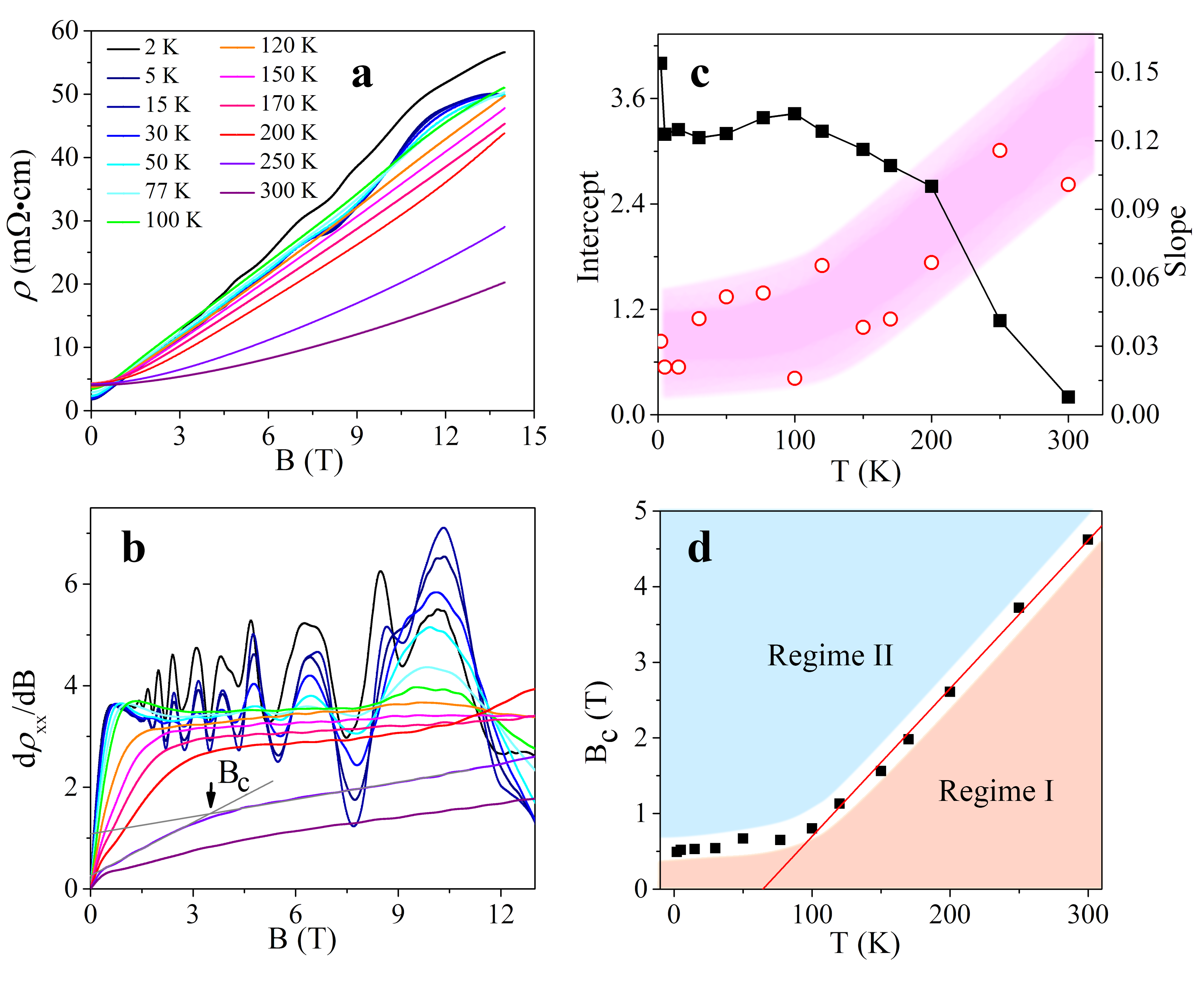}
\caption{\label{fig:fig3} {(color online) \textbf{a}, The field dependence of $\rho_{\rm xx}$ measured at different temperatures. \textbf{b}, The d$\rho_{\rm xx}$/d$B$ at different temperatures. The arrow indicates the crossover field $B_{\rm c}$ between regime I and regime II (see the context). \textbf{c}, The temperature dependencies of the intercept and the slope of d$\rho_{\rm xx}$/d$B$ in regime II. \textbf{d}, The temperature dependence of the crossover field $B_{\rm c}$.}}
\end{figure}

Figure 3\textbf{a} shows the field dependence of $\rho_{\rm xx}$ measured at different temperatures. $\rho_{\rm xx}$ changes from a quadratic field dependence near room temperatures to a linear field dependence below 100 K. This evolution can be more clearly seen from the first-order derivative of the data shown in Fig. 3\textbf{b}. The d$\rho_{\rm xx}$/d$B$ starts with a linear behavior from $B$=0, indicating that a quadratic magnetoresistive behavior dominates in the low-field regime (regime I). With increasing the magnetic field to above a crossover field $B_{\rm c}$, d$\rho_{\rm xx}$/d$B$ enters into regime II. In this regime d$\rho_{\rm xx}$/d$B$ eventually becomes flat at low temperatures, demonstrating a dominant LMR behavior.

The temperature dependencies of the intercept and the slope of d$\rho_{\rm xx}$/d$B$ in regime II are plotted in Fig. 3\textbf{c}. It can be seen that the intercept (corresponding to the amplitude of the LMR) is the biggst and almost temperature-independent below $\sim$100 K. It decreases with increasing temperature above $\sim$100 K, and finally vanishes around room temperature. Also plotted in Fig. 3\textbf{c} is the slope (corresponding to the amplitude of the quadratic component). It gradually increases with temperature, but its contribution to the total magnetoresistance is small and hardly recognizable except when the intercept drops to zero near room temperature.

The temperature dependence of the crossover field $B_{\rm c}$ is shown in Fig. 3\textbf{d}. It defines the boundary between regime I and regime II defined in Fig. 3\textbf{b}. $B_{\rm c}$ is linearly proportional to temperature above $\sim$100 K. It probably indicates that the crossover is caused by the competition between the thermal energy and a Zeeman-type of energy, i.e., $k_{\rm B}T/2$ vs. $g^*\mu_{\rm B}B_{\rm c}$, where $k_B$ is the Boltzmann constant, $g^*$ is an effective $g$ factor and $\mu_{\rm B}$ is the Bohr magneton. From the slope of the boundary we have $g^*$$\approx$40, in good agreement with the value estimated from previous study \cite{gfactor}.

\begin{figure}
\includegraphics[width=0.99 \linewidth]{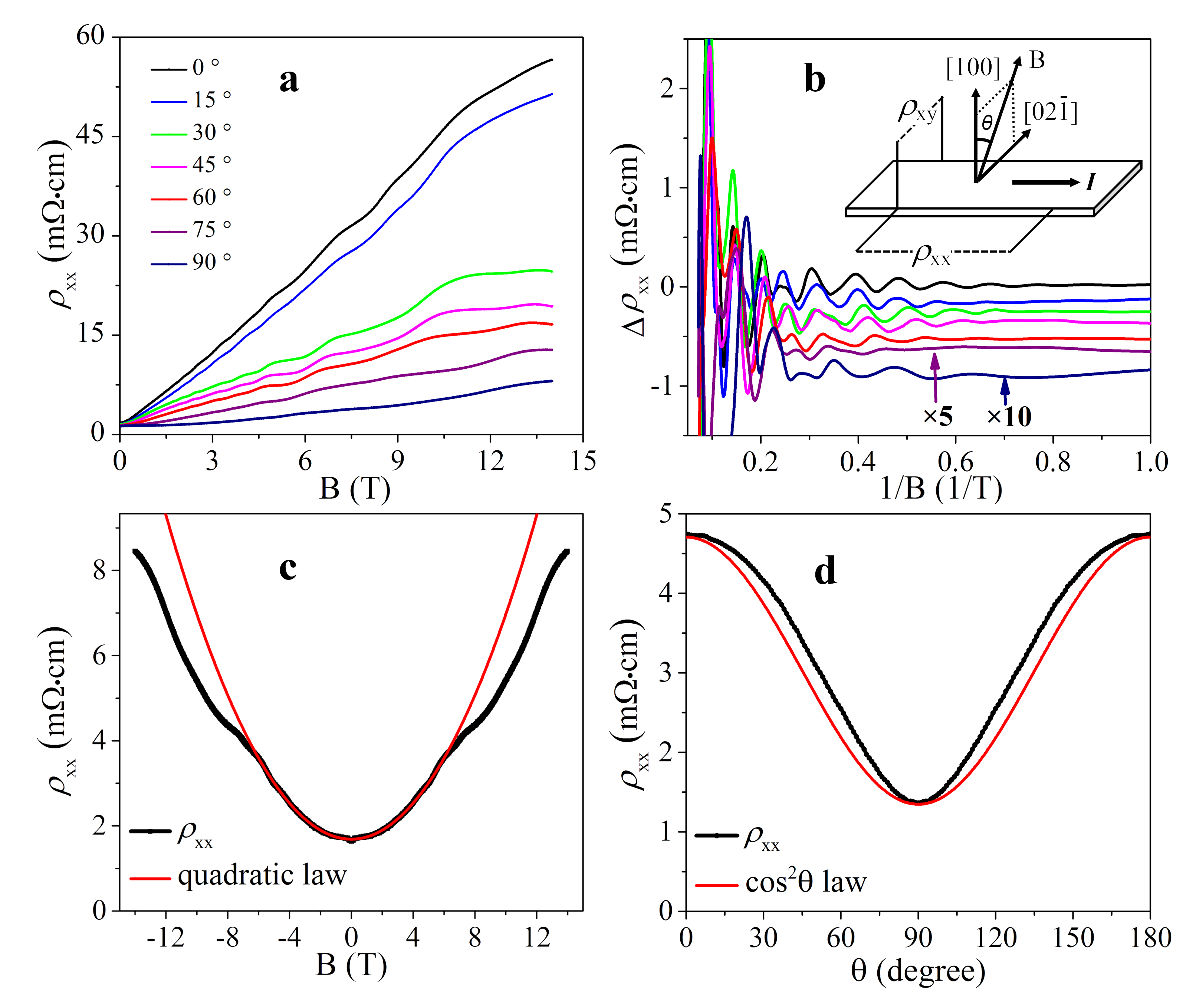}
\caption{\label{fig:fig4} {(color online) \textbf{a}, $\rho_{\rm xx}$ measured at $T$=2 K and in tilted magnetic fields, with the measurement configuration illustrated in the inset of \textbf{b}. \textbf{b}, The data in \textbf{a} are replotted against $1/B$ after the linear background of each curve is subtracted, showing the Shubnikov-de Haas oscillations. \textbf{c} $\rho_{\rm xx}$ measured at 2 K and in a field along the [02$\bar{1}]$ direction. The red line represents a quadratic law. \textbf{d}. The angular dependence of $\rho_{\rm xx}$ measured at 2 K and in 1 T. The red line is a best fit to $\cos^{2}\theta$.}}
\end{figure}

The magnetoresistive behavior varies when the magnetic field is tilted from the [100] direction to the [02$\bar{1}$] direction. The measurement configuration is illustrated in the inset of Fig. 4\textbf{b}. While $\rho_{\rm xx}$ is mostly linear in $B$ when $\theta =0^{\circ}$, it gets smaller and more quadratic as $\theta$ approaches to 90$^{\circ}$. At $\theta = 90^{\circ}$, $\rho_{\rm xx}$ can be well fitted to a quadratic law below 6 T, as shown in Fig. 4\textbf{c}.

The tilting-angle dependence of $\rho_{\rm xx}$ was also studied by continuously rotating the sample in a field of 1 T at $T$=2 K. Shown in Fig. 4\textbf{d} is the net $\rho_{\rm xx}$ subtracted from the originally measured data, based on the assumptions that $\rho_{\rm xx}(90^{\circ}-\theta)=\rho_{\rm xx}(90^{\circ}+\theta)$ and $\rho_{\rm xy}(90^{\circ}-\theta)=-\rho_{\rm xy}(90^{\circ}+\theta)$. The yielded $\rho_{xx}(\theta)$ can be fitted to the functional form of $\cos^{2}(\theta)$. We notice that similar magnetoresistance anisotropy was also observed by Liang {\it et al.}, where the field was tilted from [112] to [1$\bar{1}$0] direction, with the current applied along the [1$\bar{1}$0] direction \cite{cdasOng}.

To summarize our findings, positive and large LMR up to 3100\% in a field of 14 T was observed on Cd$_3$As$_2$ of carrier density as low as $5.0\times 10^{16} \rm cm^{-3}$. This phenomenon occurs in the measurement configuration that the current is applied along the [012] direction and magnetic field applied along the [100] direction, and when the temperature is low or magnetic field is high so that the Zeeman splitting overwhelms the thermal smearing. The linear behavior gradually degrades to a quadratic one when the magnetic field is tilted from the [100] direction to the [02$\bar{1}$] direction of the plate-like crystal.

To explain the LMR behavior, several mechanisms mainly in two classes have been proposed. One believes that the LMR arises by picking up the Hall signal in inhomogeneous samples \cite{Littlewood03}. In this picture, the sample is modeled as a network due to the existence of large mobility fluctuation. The low carrier density yields a large Hall signal on the network which eventually influences the $\rho_{\rm xx}$ measurement. This picture, used by Parish and Littlewood to explain the LMR in highly disordered Ag$_{2\pm\delta}$Se and Ag$_{2\pm\delta}$Te, is not likely applicable for our high-quality Cd$_{3}$As$_{2}$ single crystals.

In another class of mechanisms, the interpretation of LMR relies on the formation of LLs. Abrikosov proposed that a linear quantum magnetoresistance arise when all the electrons are filled into the first LL \cite{Abrikosov98, Abrikosov2000}. In this picture, the amplitude of LMR is independent of mobility but only depending on the carrier density. These features seem to fit with our data, i.e., although the mobility changes roughly by a factor of 2 below $\sim$100 K, the carrier density keeps almost constant (Fig. 2\textbf{d}), so does the gross slope of LMR (Fig. 3\textbf{a}). However, the Abrikosov mechanism requires that all the electrons are filled into the first LL, which is not the case in our experiment.

Magnetoresistance in ordinary metals is usually much smaller than the resistance itself. A LMR up to 3100\% in a magnetic field of 14 T must be caused by some peculiar mechanisms, in which either the electron density of states, or the scattering rate, or both, must be significantly modified by applying a magnetic field. Here the modification in the electron density of states refers to the modification in the Fermi surface. And the modification in the scattering rate (hence the carrier mobility) refers to the change of scattering processes which are otherwise prohibited in zero magnetic field.

Among various possible mechanisms, here we would draw attention to the field-induced peculiar distortion/shifting of the helical electrons' Fermi surface in momentum space. Due to their opposite helicities, the two Weyl Fermi surfaces of each 3D Dirac cone, which are originally overlapped in zero magnetic field, will be distorted and shifted apart in the momentum space by applying a magnetic field. This process is more significant in samples with lower carrier density, such that the Fermi surfaces are small and close to the Dirac points, rather than touching with each other (the Lifshitz transition). The field-induced distortion/shifting would lead to the change in electron density of states at the Fermi energy, and would also allow the happening of backscattering between the two Weyl Fermi surfaces or even within each Weyl Fermi surface. From the data of SdH oscillations shown in Figs. 3\textbf{b} and 4\textbf{b}, in which the oscillation period keeps roughly constant over the entire $1/B$ range, we feel that the field-induced changes in electron density of states is insufficient to account for the observed LMR up to 3100\%. We therefore believe that the field-induced distortion/shifting of the Fermi surfaces would release the backscattering processes and giving rise to the LMR, after the Zeeman splitting overwhelms the thermal smearing. We note that lifting the protection of backscattering as a possible mechanism of LMR has also been indicated in a recent experiment by Liang and coworkers \cite{cdasOng}.

The above scenario also explains why a large LMR is often observable in Dirac electron systems. Dirac electron systems possess all the key elements required in such a scenario, i.e., containing helical electrons, having small Fermi surfaces, yet with large Zeeman splitting due to large $g$ factor. For Cd$_3$As$_2$, its $g^*$$\approx$40, the Zeeman splitting is $\sim$23 meV at $B$=10 T. This energy scale is comparable to the Fermi energy of $\sim$30 meV (SdH) or $\sim$40 meV (Hall) estimated from the carrier density of the sample (explained the supplementary materials).

In the above scenario, however, the anisotropy of the LMR remains unexplained. Further theoretical and experimental studies are needed to test this scenario, and to fully clarify the physics behind the LMR behavior.

\vspace {0.3 cm}

\begin{acknowledgments}
We would like to thank C. L. Yang for helpful discussions. This work was supported by the National Basic Research Program of China from the MOST under the Contracts No. 2009CB929101, No. 2011CB921702 and No. 2011CBA00110, by the NSFC under Contracts No. 91221203, No. 11174340, No. 11174357 and No. 11274367, and by the Knowledge Innovation Project and the Instrument Developing Project of CAS.
\end{acknowledgments}

\vspace {0.5 cm}



\section*{Supplementary material}

# Supplementary materials for "Large linear magnetoresistance in Driac semi-metal $Cd_3As_2$ with Fermi surfaces close to the Dirac points"

Junya Feng, Yuan Pang, Desheng Wu, Zhijun Wang, Hongming Weng, Jianqi Li, Xi Dai, Zhong Fang, Youguo Shi,* and Li Lu†
*Beijing National Laboratory for Condensed Matter Physics,*
*Institute of Physics, Chinese Academy of Sciences*
*Collaborative Innovation Center of Quantum Matter, Beijing 100190, People's Republic of China*
(Dated: May 26, 2014)

## CRYSTAL SYNTHESIS

Single crystals of $Cd_3As_2$ were synthesized by a chemical vapor transport technique with two steps. First, the starting materials of cadmium shot (99.999%) and arsenic lump (99.9999%) were mixed in stoichiometric ratio of 3:2 and placed in an alumina ampoule. Then the alumina ampoule which contains the mixtures was sealed in a quartz tube under Argon atmosphere. The mixture was heated to 750°C at a rate of 3°C/min and kept for 20 hours in a furnace. Then we cooled the quartz tube down to room temperature. Second, the quartz tube was transferred into a tube furnace to make a temperature gradient. The length of the quartz tube used for vapor transition was about 15 cm. One end (the high temperature area) of the quartz tube which contains the compound was heated up to 690°C at a rate of 3°C/min, while the low temperature part was less than 580°C. The quartz tube was kept in the tube furnace for one week then cooled down to room temperature at a rate of 5°C/min. Finally, many shiny mirror-like crystals were obtained.

Phase identification of the crystals was performed at room temperature on a Shimadzu XRD-7000 X-ray diffractometer using Cu K$\alpha$1($\lambda = 1.5406$Å). XRD data were collected in a $2\theta$ range of $5° - 80°$ with a scan step interval of $0.02°$.

## THE HEAT CAPACITY

Figure S1 is the temperature dependence of the heat capacity of $Cd_3As_2$ crystals. It shows that there is no structural phase transition in the temperature range of 2 K to room temperature.

## THE CARRIER DENSITY ESTIAMTED FROM SDH OSCILLATIONS

It can be found in the text books [1] that for a 3D system, the relation between carrier density and SdH oscillation period is given by the Dirac-like band structure:

$$\Delta\left(\frac{1}{B}\right) = \frac{2e}{\hbar}\left(\frac{g_s g_v}{6\pi^2 n_{3D}}\right)^{2/3} \quad (1)$$

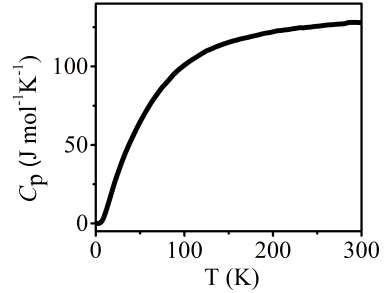

FIG. S1 The heat capacity of $Cd_3As_2$ crystals from room temperature to 2 K, showing no signature of structural phase transition.

where $g_s$ is the spin degeneracy, and $g_v$ the valley degeneracy. With $\Delta(1/B) = 0.0932$ T$^{-1}$ for the specimen presented in the main paper, we have $n_{3D} = 1.25 \times 10^{17} cm^{-3}$ if assuming $g_s g_v = 2$, i.e., assuming that the spin degeneracy has been lifted.

We can also calculate the carrier density from the Hall resistance directly. With the $R_H$ data and taking the thickness of the sample to be 20$\mu$m, we have $n_H = 4.95 \times 10^{16} cm^{-3}$.

The electron mobility can further be estimated using the relation $\rho = 1/ne\mu$. Since $\rho = 1.8$m$\Omega \cdot$cm in zero field and at 2 K, we have $\mu = 7.0 \times 10^4$ cm$^2$V$^{-1}$s$^{-1}$.

## THE EFFECTIVE MASS AND THE DINGLE TEMPERATURE

Figure S2 **a** displays that SdH oscillations at different temperatures. After the linear background of $R_{xx}$ is subtracted, the SdH oscillations of $\Delta R_{xx}$ can be clearly seen up to a temperature of 77 K. The amplitude of the SdH oscillations decreases with increasing temperature. We fitted the temperature dependence of the SdH oscillation amplitude $\Delta R_{xx}$ using the Lifshitz-Kosevich theory: [1]

$$\Delta R_{xx}(T,B) \propto \frac{2\pi^2 k_B T/\Delta E_N(B)}{\sinh[2\pi^2 k_B T/\Delta E_N(B)]} \quad (2)$$
$$\times e^{-2\pi^2 k_B T_D/\Delta E_N(B)}$$

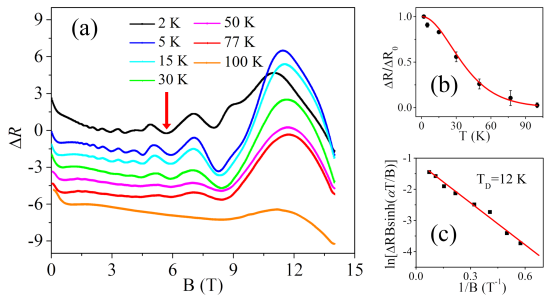
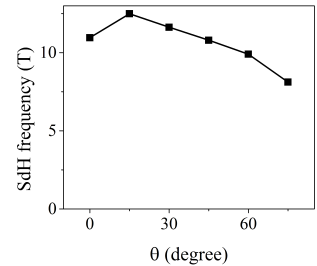

FIG. S2 (color online) **a**, Field dependence of $R_{xx}$ between 2 K and 100 K (For each curve its linear background has been subtracted, and its vertical position is shifted for clarity). **b**, Temperature dependence of the relative amplitude of SdH oscillations of $R_{xx}$ for $\theta = 0°$ at 2 K. The red line is a best fit to the Lifshitz-Kosevich formula. **c**, Dingle plot of the oscillations in $R_{xx}$, indicating a Dingle temperature of 12 K.

where $T_D = h/4\pi^2\tau k_B$ is the Dingle temperature, $\tau$ is the quantum lifetime of the carriers, $\Delta E_N(B) = heB/2\pi m^*$ is the energy gap between the $N^{th}$ and the $(N+1)^{th}$ Landau level (LL), $h$ is the Planck constant, $e$ is the electron charge, and $m^*$ is the effective mass of the carriers.

Plotted in Fig. S2**b** is the relative amplitude of oscillation $\Delta R/\Delta R_{2K}$ as a function of temperature for the LL located at the magnetic field marked by the red arrow in Fig. S2**a**. By fitting this curve, we extracted the carriers' effective mass $m^* = 0.025 m_e$ (where $m_e$ is the free electron mass). The result is consistent with previous study [5]. In Fig. S2**c**, the amplitudes of the SdH oscillations at 2 K are used in a $\ln[\Delta R B \sinh(\alpha T/B)]$ vs $1/B$ plot. From the slope, $T_D = 12$ K can be determined. The quantum lifetime of the carriers can further be calculated: $\tau = 1.05 \times 10^{-13}$ s.

## ANGULAR DEPENDENCE OF THE FERMI SURFACE

Figure S3 shows the angular dependence of the SdH oscillation frequency of the specimen studied in the main paper.

## THEORETICAL ANALYSIS ON THE CARRIER DENSITY, FERMI ENERGY AND THE LIFSHITZ TRANSITION

The relation between the carrier density and the Fermi energy of $Cd_3As_2$ is obtained via numerical simuation, and the results are shown in Fig. S4. From the measured carrier density, the Fermi energy of the $Cd_3As_2$ specimen investigated in this experiment is $\sim 30$ meV (using the carrier density of the sample estimated from the S-dH oscillations) or $\sim 40$ meV (using the carrier density estimated from the Hall data).

FIG. S3 (color online) The angular dependence of the SdH oscillation frequency of the specimen studied in the main paper.

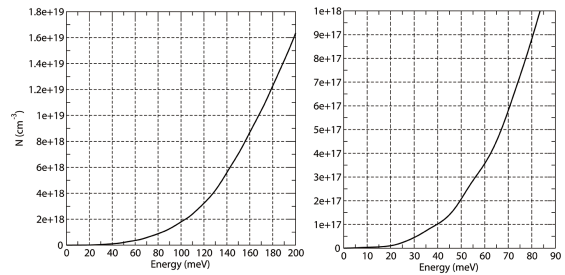

FIG. S4 (color online) The calculated relation between the carrier density and the Fermi energy of $Cd_3As_2$. The right panel magnifies the low energy part of the left one.

Figure S5 shows the calculated Fermi surface of $Cd_3As_2$ with different Fermi energy, namely 50 meV (**a**), 100 meV (**b**), 133 meV (**c**) and 150 meV (**d**). Lifshitz transition happens when the two Fermi surface touches with each other when $E_F = 133$ meV (Fig. S5**c**). The Fermi surface of single $Cd_3As_2$ crystal we used is around or smaller than $\sim 40$ meV, well below the Lifshitz transition.

---

* Corresponding author for crystal growth: ygshi@iphy.ac.cn
† Corresponding author: lilu@iphy.ac.cn

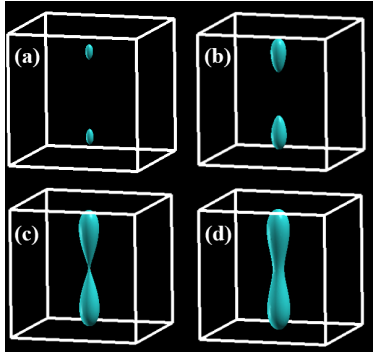

FIG. S5 (color online) The calculated Fermi surface of $Cd_3As_2$ at several different Fermi energies, namely 50 meV **(a)**, 100 meV **(b)**, 133 meV **(c)** and 150 meV **(d)**.

99



\end{document}